\documentclass[12pt]{article}
\usepackage[T1]{fontenc}
\usepackage{amsfonts}
\usepackage{latexsym}
\usepackage{dsfont}
\usepackage{amsmath}
\usepackage{amssymb}
\usepackage{amssymb}
\usepackage{amsthm}
\usepackage{mathrsfs}
\usepackage[nosort]{cite}
\usepackage[bulletsep]{collref}
\usepackage{setspace}
\usepackage{color}
\usepackage{graphicx}
\usepackage{accents}
\usepackage[font=small,labelfont=bf]{caption}
\usepackage[utf8]{inputenc}
\usepackage{geometry}
\usepackage[toc,page]{appendix}

\usepackage{tikz}
\usetikzlibrary{arrows.meta,backgrounds,calc,fit,positioning,scopes,shadows}

\makeatletter
\def\tikzsavelastnodename#1{\let#1=\tikz@last@fig@name}
\makeatother

\usepackage[hidelinks]{hyperref}
\hypersetup{
colorlinks=false,
citecolor= blue,
linkcolor= blue,
urlcolor= blue,
breaklinks=true
}
\usepackage{cleveref}
\crefformat{section}{\S#2#1#3} 
\crefformat{subsection}{\S#2#1#3}
\crefformat{subsubsection}{\S#2#1#3}

\def\appendix#1{\addtocounter{section}{1}\setcounter{equation}{0}
\renewcommand{\thesection}{\Alph{section}}
\section*{Appendix \thesection\protect\indent \parbox[t]{11.15cm}{#1}}
\addcontentsline{toc}{section}{Appendix \thesection\ \ \ #1}}

\allowdisplaybreaks
\numberwithin{equation}{section}

\makeatletter
 \let\old@startsection=\@startsection
 \let\oldl@section=\l@section
 \renewcommand{\@startsection}[6]{\old@startsection{#1}{#2}{#3}{#4}{#5}{#6\mathversion{bold}}}
 \renewcommand{\l@section}[2]{\oldl@section{\mathversion{bold}#1}{#2}}
\makeatother

\def\dd{\text{d}}

\DeclareMathOperator*{\arcsinh}{arcsinh}
\DeclareMathOperator*{\arctanh}{arctanh}

\begin{document}


\begin{titlepage}
\begin{center}
\vspace*{-1.0cm}
\hfill {\footnotesize ZMP-HH/23-9}

\vspace{2.0cm}

{\LARGE  {\fontfamily{lmodern}\selectfont \bf Flowing from relativistic to \\
\vspace{2mm}
non-relativistic string vacua in AdS$_5 \times$S$^5$}} \\[.2cm]

\vskip 1.5cm
\textsc{Andrea Fontanella$^{\mathfrak{Re}}$ \footnotesize and \normalsize Juan Miguel Nieto Garc\'ia$^{\mathfrak{Im}}$}\\
\vskip 1.2cm

\begin{small}
{}$^{\mathfrak{Re}}$ \textit{Perimeter Institute for Theoretical Physics, \\
Waterloo, Ontario, N2L 2Y5, Canada} \\
\vspace{1mm}
\href{mailto:afontanella@perimeterinstitute.ca}{\texttt{afontanella@perimeterinstitute.ca}}

\vspace{5mm}
{}$^{\mathfrak{Im}}$ \textit{II. Institut für Theoretische Physik, Universität Hamburg,\\
Luruper Chaussee 149, 22761 Hamburg, Germany} \\
\vspace{1mm}
\href{mailto:juan.miguel.nieto.garcia@desy.de}{\texttt{juan.miguel.nieto.garcia@desy.de}}

\end{small}

\end{center}

\vskip 1 cm
\begin{abstract}
\vskip1cm\noindent
We find the connection between relativistic and non-relativistic string vacua in AdS$_5 \times$S$^5$ in terms of a free parameter $c$ flow. First, we show that the famous relativistic BMN vacuum flows in the large $c$ parameter to an unphysical solution of the non-relativistic theory.
Then, we consider the simplest non-relativistic vacuum, found in \href{https://arxiv.org/abs/2109.13240}{arXiv:2109.13240} (called BMN-like), and we identify its relativistic origin, namely a non-compact version of the folded string with zero spin, ignored in the past due to its infinite energy. We show that, once the critical closed B-field required by the non-relativistic limit is included, the total energy of such relativistic solution is finite, and in the large $c$ parameter it precisely matches the one of the BMN-like string. We also analyse the case with spin in the transverse AdS directions.  
\end{abstract}

\end{titlepage}

\tableofcontents
\vspace{5mm}
\hrule


\setcounter{section}{0}
\setcounter{footnote}{0}

\section*{Introduction}

The problem of finding the string spectrum is in general a difficult task. When the background is flat space, the spectrum of fluctuations is independent of the classical solution around which one expands the action. The reason behind it is that flat space string sigma model is Gaussian. This is not the case when the background is a curved manifold, such as AdS$_5 \times$S$^5$, where the spectrum will in general depend on the particular classical string solution chosen for the action expansion. Regarding relativistic strings in AdS$_5 \times$S$^5$, the problem of computing the spectrum around the BMN vacuum has been well studied with many integrability techniques (Thermodynamic Bethe ansatz, Y-system, etc., see \cite{Beisert:2010jr} for a review on the topic), but the spectrum for other vacua is a more complicated problem, and to the best of our knowledge, the only other vacuum where integrability techniques have been applied is the GKP vacuum \cite{Basso:2010in,Fioravanti:2011xw,Fioravanti:2013eia}. In recent years, there are indications that the problem of finding the spectrum around a general vacua can be solved (at least, formally) using an integrability-based technique called Quantum Spectral Curve, see \cite{Levkovich-Maslyuk:2019awk} for a review on the topic. The success in understanding strings in flat space and AdS$_5 \times$S$^5$ has motivated a renewed interest in theories derived from or reminiscent of them, like $T\bar{T}$ deformations, Yang-Baxter deformations, fishnet CFT, lower-dimensional AdS, or non-relativistic limit.

In this article, we focus on the non-relativistic (NR) limit of string theory. We consider a string theory where the NR limit has been taken on the target space geometry, whereas the world-sheet remains relativistic. In this limit, the background geometry probed by the string is a String Newton-Cartan (SNC) geometry, namely a particular non-Lorentzian type of geometry.  In this setting, bosonic NR string theory defined on a generic SNC target space is free of Weyl anomalies provided the beta function vanishes \cite{Gomis:2019zyu,Gallegos:2019icg}. 

The first example of NR string theory was proposed in flat space in \cite{Gomis:2000bd, Danielsson:2000gi}, and the only known example so far of NR string in curved background is the one found in \cite{Gomis:2005pg}, which is formulated by taking a NR limit in the AdS$_5 \times$S$^5$ geometry while keeping the world-sheet relativistic. It is a current open problem to explore the landscape of NR string theory, as there are indications that many backgrounds, which are different at the relativistic level, will all limit to the same type of background proposed in \cite{Gomis:2000bd, Gomis:2005pg}, suggesting that the landscape of NR strings consists only in the flat space and AdS$_5 \times$S$^5$ SNC geometries.

NR string theory has been extensively explored at the formal level. The structure of the NR Polyakov string action in any SNC background can be obtained from several different but equivalent approaches. A first approach, described in \cite{Bergshoeff:2018yvt} and known as the limit procedure, involves rescaling the target space vielbein by a parameter that is then sent to infinity. A second approach, is the so-called null reduction approach \cite{Harmark:2017rpg, Harmark:2018cdl, Harmark:2019upf}, which consists in reducing a relativistic string theory on a Lorentzian manifold along a null isometry at fixed momentum. A third approach, proposed in \cite{Hartong:2021ekg, Hartong:2022dsx}, consists in taking an expansion of the relativistic action in a large parameter $c$ and study the equations of motion up to a certain order in $c$. The idea behind the expansion approach is reminiscent of the Lie algebra expansion method applied to coset string sigma models \cite{Fontanella:2020eje}, where in the latter approach it is required a concrete target space isometry algebra in order to apply the expansion on it. Related topics, such as T-duality \cite{Bergshoeff:2018yvt}, symmetries of the action \cite{Harmark:2019upf,  Bergshoeff:2019pij, Bidussi:2021ujm}, connection to double field theory \cite{Ko:2015rha, Blair:2019qwi, Blair:2020gng, Blair:2020ops}, Hamiltonian formalism \cite{Kluson:2017abm,Kluson:2018egd,Kluson:2018grx}, open strings \cite{Gomis:2020fui,Gomis:2020izd}, NR supergravity \cite{Bergshoeff:2021bmc, Bergshoeff:2021tfn} and theories with NR world-sheet \cite{Harmark:2017rpg, Harmark:2018cdl, Harmark:2019upf,Harmark:2017yrv, Harmark:2020vll, Baiguera:2020jgy, Baiguera:2020mgk, Baiguera:2021hky} have also been studied. We refer to \cite{Oling:2022fft} and references therein for a recent review on the topic.

Although much progress has been achieved from the formal point of view of NR string theory, less is known about its physics, in particular regarding its predictions, such as how observables look like in this theory. 
Having knowledge on this would be particularly appealing in view of formulating a NR version of holography, which would be of non-AdS type as the target space is a SNC manifold. 

As at the moment the best understood example of holography is given by relativistic string theory in AdS$_5 \times$S$^5$, it seems reasonable to explore what is its NR version.
This would amount to formulating a holographic correspondence between NR string theory in SNC AdS$_5 \times$S$^5$, e.g. \cite{Gomis:2005pg}, and its, not yet known, dual field theory.  Although we still have not yet identified the dual field theory, some progress on the string theory side has been made. In particular, a coset description of the SNC AdS$_5 \times$S$^5$ manifold was found \cite{Fontanella:2022fjd, Fontanella:2022pbm}, which made possible to formulate a NR version of the Metsaev-Tseytlin coset action. Thanks to this formalism, it was possible to find a Lax pair for this type of theory \cite{Fontanella:2022fjd}, which is the first step towards showing classical integrability.   

In view of a NR version of holography, it would be important to determine the full string spectrum of NR string theory in SNC AdS$_5 \times$S$^5$.\footnote{In the flat space case, the NR string spectrum can be easily accessed by taking the zero Regge limit of the relativistic spectrum \cite{Gomis:2000bd}. Such procedure is vacuum independent, as the flat space action is Gaussian.} Although a final answer to this problem is still missing, some progress in this direction has been made. Classical string solutions admitted by NR string theory in SNC AdS$_5 \times$S$^5$ were studied in \cite{Fontanella:2021btt}, where it was shown that the simplest vacuum admitted by this theory must always have a winding along the longitudinal direction, as demanded by  consistency with the Lagrange multipliers equations of motion. Such physical requirement renders even the simplest vacuum, which was called BMN-like, quite complicated when expanding the action around it \cite{Fontanella:2021hcb}, spoiling the usual perturbative S-matrix computation.  
A different approach to the spectrum in NR string theory in SNC AdS$_5 \times$S$^5$ was given by analysing the classical spectral curve associated with the Lax pair evaluated on the BMN-like vacuum \cite{Fontanella:2022wfj}. The non-standard result obtained indicates that the usual notion of spectral curve should be generalised to take into account the non-diagonalisability of the monodromy matrix in the context of non semi-simple isometry algebras, which is the case when taking the NR limit.

The motivation of this article is to explore whether it is sensible in AdS$_5 \times$S$^5$ to construct the spectrum above NR classical strings by taking directly the limit of the relativistic string spectrum, similarly to the flat space case, instead of having to construct it from scratch using the NR action. For that, we shall show that the profile and the conserved charges of NR classical strings can indeed be obtained as a limit of an appropriate relativistic classical string.

Our strategy consists in considering relativistic strings in AdS$_5 \times$S$^5$ where we keep the NR rescaling parameter $c$. When $c=1$, we get the usual relativistic action and when $c\rightarrow \infty$, we get NR string theory in SNC AdS$_5 \times$S$^5$. This flowing\footnote{When we use the word \emph{flow} we mean that the formulas involved in our study have a parametric dependence on a free parameter $c$ which goes from 1 to $\infty$ .} procedure requires to couple the action to a critical closed B-field, in order to correctly treat the divergent term appearing in the action when $c\rightarrow \infty$. By keeping the $c$ parameter free, we are able to identify the classical solution of the relativistic theory, which flows at large $c$ to the BMN-like vacuum of NR string theory in SNC AdS$_5 \times$S$^5$. Such solution is a non-compact version of the folded string with zero spin, and it has infinite energy. However, once the critical B-field is considered, we find that the divergent part of the energy cancels out exactly. The total energy, which is the sum of the contributions from the metric and the critical B-field, remains finite in the whole flow and, when $c\rightarrow \infty$, it reproduces precisely the energy of the BMN-like solution. 

We also study the case when the solution has a spin in the transverse AdS directions. In this case, we will find that the total energy and spin are both divergent. However, for a precise fine-tuning of the free parameters of the solution, we show that in the large $c$ limit a linear combination of the total energy and the spin is finite and precisely matches the dispersion relation and profile of a spinning solution found by solving the equations of motion in the NR string theory in SNC AdS$_5 \times$S$^5$. 

\begin{figure}[t]
\begin{center}
\includegraphics[width=\textwidth,keepaspectratio]{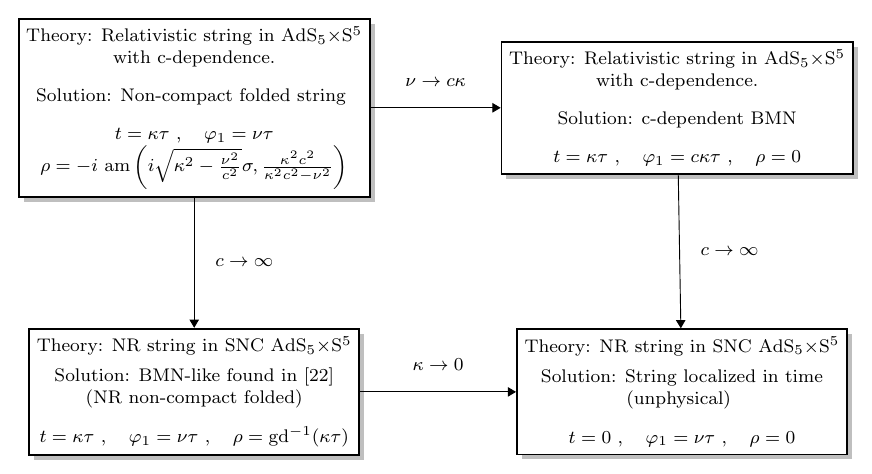}
    
\end{center}

\caption{This diagram illustrates the relativistic origin of the BMN-like string of NR string theory in SNC AdS$_5 \times$S$^5$. It also shows that the relativistic BMN vacuum is not a consistent vacuum for such theory.}
    \label{fig:flow}
\end{figure}

Another important point is to consider the flow of the relativistic BMN string. We show that in the large $c$ limit this solution becomes an unphysical string localised in time. This is an evidence that the BMN vacuum, which is perhaps the simplest vacuum for relativistic strings in AdS$_5 \times$S$^5$, does not survive when $c$ is taken to be large. This can be interpreted as the fact the BMN vacuum describes a classical point-like string fast moving around the equator of the 5-sphere and therefore such motion cannot be seen at the slow velocity regime captured by the NR string theory in SNC AdS$_5 \times$S$^5$.
A diagram summarising the flow in the case of zero spin is given in Figure \ref{fig:flow}.

This article is organised as follows. In section \ref{sec:NR_solutions} we review how NR string theory in SNC AdS$_5 \times$S$^5$ is obtained in the limit procedure from relativistic string action in AdS$_5 \times$S$^5$. We present solutions parameterised by three parameters $(\kappa, \omega, \nu)$, associated with energy, spin and angular momentum, which are the polar coordinate version of solutions found in \cite{Fontanella:2021btt}. In section \ref{sec:flow_rel} we solve the equations of motion for the relativistic string action in AdS$_5 \times$S$^5$ keeping the free parameter $c$. We present three solutions, namely the BMN, a non-compact version of the folded string with zero spin, characterised by the parameters $(\kappa, 0, \nu)$, and its spinning version, with parameters $(\kappa, \omega, \nu)$. We study their Noether charges, dispersion relation and large $c$ limit. In section \ref{sec:conclusions} we give our concluding summary and future prospects.  
The article ends with two appendices, one on the convention and the other one on the NR rescaling in polar coordinates.

\section{Non-relativistic string action and classical solutions}
\label{sec:NR_solutions}

In this section, we review the construction of the bosonic sector of NR string theory in SNC AdS$_5 \times$S$^5$ starting from the bosonic sector of relativistic AdS$_5 \times$S$^5$ string action. We construct the simplest classical solutions admitted by NR string theory in SNC AdS$_5 \times$S$^5$ and their spinning generalisation, which are the analogue in polar coordinates of the solutions constructed in Cartesian coordinates in \cite{Fontanella:2021btt}.

\subsection{Relativistic and non-relativistic actions}

 For the purpose of studying classical string solutions where fermion fields vanish, we consider the bosonic sector of type IIB superstring theory in AdS$_5 \times$S$^5$ given by
\begin{equation}
\label{rel_action}
    S=-\frac{\sqrt{\lambda}}{4\pi}  \int{\dd^2 \sigma \bigg( \gamma^{\alpha \beta} \partial_\alpha X^\mu \partial_\beta X^\nu g_{\mu \nu} +  \varepsilon^{\alpha\beta} \partial_\alpha X^\mu \partial_\beta X^\nu b_{\mu \nu} \bigg)},
\end{equation}
where $\lambda$ is the `t Hooft coupling, related to the string tension $T$ via $2\pi T = \sqrt{\lambda}$, the string world-sheet coordinates are collected as $\sigma^{\alpha} = (\tau, \sigma)$, with $\sigma \equiv \sigma + 2 \pi$, and $\gamma^{\alpha\beta} \equiv \sqrt{-h} h^{\alpha\beta}$ is the Weyl invariant combination of the inverse worldsheet metric $h^{\alpha\beta}$ and $h =$ det$(h_{\alpha\beta})$. 
The metric $g_{\mu\nu}$ is the AdS$_5 \times$S$^5$, 
\begin{align}
    \dd s^2 &=g_{\mu \nu} dX^\mu dX^\nu=    \dd s^2_{AdS}+\dd s^2_{S}  \ , \label{coord}\\
    \notag
    \dd s^2_{AdS} &= -\cosh^2 \rho \, \dd t^2 +\dd\rho^2 + \sinh^2 \rho \, \dd\beta_1^2 + \sinh^2 \rho \, \cos^2 \beta_1 \, \dd\beta_2^2 \\
    \notag
    &+ \sinh^2 \rho \, \cos^2 \beta_1 \, \cos^2 \beta_2 \, \dd\beta_3^2 \ , \notag \\
    \notag
    \dd s^2_{S} &= \dd \varphi_1^2 + \cos^2 \varphi_1 \, \dd \varphi_2^2 + \cos^2 \varphi_1 \, \cos^2 \varphi_2 \, \dd\varphi_3^2 + \cos^2 \varphi_1 \, \cos^2 \varphi_2 \, \cos^2 \varphi_3 \, \dd\varphi_4^2  \\
    &+\cos^2 \varphi_1 \, \cos^2 \varphi_2 \, \cos^2 \varphi_3 \, \cos^2 \varphi_4 \, \dd\varphi_5^2\ .  \notag 
\end{align}
Here $t$ is the global time, $\rho$ is the radial coordinate in AdS$_5$ and $\beta_i$ and $\varphi_j$ are angles in AdS$_5$ and S$_5$ respectively. The B-field $b_{\mu\nu}$ is a closed Kalb-Ramond field, i.e. $\dd b=0$, which we will fine tune to a critical value when taking the NR limit.

The analogue in polar coordinates of the NR limit considered in \cite{Gomis:2005pg} is given by rescaling the coordinates (see Appendix \ref{App:polar_coset} for details)
\begin{equation}
\label{coord_resc}
\beta_2 + \frac{\pi}{2} \rightarrow \frac{1}{c} \big(\beta_2 + \frac{\pi}{2} \big) \ , \qquad
\beta_1 \rightarrow \frac{1}{c} \beta_1 \ , \qquad
\varphi_2 + \frac{\pi}{2} \rightarrow \frac{1}{c} \big(\varphi_2 + \frac{\pi}{2} \big) \ , \qquad
\varphi_1 \rightarrow \frac{1}{c} \varphi_1 \ ,
\end{equation}
while leaving the other coordinates invariant. In addition, we also have to rescale the string tension as $T \rightarrow c^2 T$, which can be absorbed into a more convenient redefinition of the metric and B-field, 
\begin{eqnarray}
    G_{\mu\nu} \equiv c^2 g_{\mu\nu} \ , \qquad\qquad
    B_{\mu\nu} \equiv c^2 b_{\mu\nu} \ .
\end{eqnarray}
Here $c$ is a non-relativistic contraction parameter, which plays the stringy analogue role of the speed of light, and is assumed to be large. 
At large values of $c$, the vielbein associated to $G_{\mu\nu}$ expands as
\begin{align}
 \hat{E}_{\mu}{}^A &= c \tau_{\mu}{}^A + \frac{1}{c} m_{\mu}{}^A + \mathcal{O}(c^{-3})\ , &
 \hat{E}_{\mu}{}^a &= e_{\mu}{}^a + \mathcal{O}(c^{-2}) \ , \label{vielbeine_exp}
\end{align}
where $A= 0 , 1$ (longitudinal) and  $a = 2, ..., 9$ (transverse). The set of vielbein $\{\tau_{\mu}{}^A, m_{\mu}{}^A, e_{\mu}{}^a\}$ are called Newton-Cartan data, and in polar coordinates are\footnote{Here we trade some abuse of notation for some convenience by writing $\tau_\mu{}^A$ and $e_\mu{}^a$ as $10\times 10$ matrices, although they are $10\times 2$ and $10 \times 8$ matrices respectively. In this matrix representation, the $\mu$ index runs over coordinates in the same order in which their differentials appears in (\ref{coord}).}
\begin{align}
    \tau_\mu{}^A =&\text{diag} \left( \cosh \rho , 1 , 0 , 0 ,0 ,0 ,0 ,0 ,0 ,0 \right) \ , \qquad\qquad\qquad m_\mu{}^A = 0 \ , \\
    e_\mu{}^a =&\text{diag} \left( 0 , 0 , -\sinh \rho , -\sinh \rho , \sinh \rho \, (\beta_2 + \pi/2) , 1 , 1 , \right. \notag \\
    &\left.-(\varphi_2 - \pi/2) , -(\varphi_2 - \pi/2) \, \cos \varphi_3 , (\varphi_2 - \pi/2) \, \cos \varphi_3 \, \cos \varphi_4 \right) \ .
\end{align}
Notice that $\tau_\mu{}^A$ is an AdS$_2$ vielbein and $e_\mu{}^a$ is the vielbein of (sinh$\rho\,  \mathbb{R}^3 ) \times \mathbb{R}^5$.  Substituting this expansion into the action gives us
\begin{equation}
\label{S_NR_div}
S = - \frac{\sqrt{\lambda}}{4\pi}  \int \dd^2 \sigma \, \gamma^{\alpha\beta} \bigg( c^2 \partial_{\alpha} X^{\mu} \partial_{\beta} X^{\nu} \tau_{\mu\nu} + \partial_{\alpha} X^{\mu} \partial_{\beta} X^{\nu} H_{\mu\nu} + \varepsilon^{\alpha\beta}  \partial_{\alpha} X^{\mu} \partial_{\beta} X^{\nu} B_{\mu\nu}\bigg) + \mathcal{O}(c^{-2}) \ , 
\end{equation} 
where
\begin{align}
    \tau_{\mu \nu}&\equiv \tau_{\mu}{}^A \tau_{\nu}{}^B \tilde{\eta}_{AB}= \text{diag} \left( -\cosh^2 \rho , 1 , 0 , 0 , 0 , 0 , 0 , 0 , 0 , 0 \right) \ , \\
    H_{\mu \nu}&\equiv e_{\mu}{}^a e_{\nu}{}^b \tilde{\delta}_{ab} + \bigg(\tau_{\mu}{}^A m_{\nu}{}^B + \tau_{\nu}{}^A m_{\mu}{}^B \bigg) \tilde{\eta}_{AB} \notag \\
    &=\text{diag} \left( 0 , 0 , \sinh^2 \rho , \sinh^2 \rho , \sinh^2 \rho \, (\beta_2 + \pi/2)^2 , 1 , 1 , \right. \notag \\
    &\left.(\varphi_2 - \pi/2)^2 , (\varphi_2 - \pi/2)^2 \, \cos^2 \varphi_3 , (\varphi_2 - \pi/2)^2 \, \cos^2 \varphi_3 \, \cos^2 \varphi_4 \right) \ ,
\end{align}
with $\tilde{\eta}_{AB} =$ diag$(-1, 1, 0, ..., 0)$, $\tilde{\delta}_{ab} =$ diag$(0,0,1,...,1)$.  Although the action diverges for large values of $c$, we can combine the divergent term with a fine-tuned B-field of the form
 \begin{equation}
B \equiv \frac{c^2}{2} \tau_{\mu}{}^A \tau_{\nu}{}^B \varepsilon_{AB} \, \dd X^{\mu} \wedge \dd X^{\nu} = - \frac{c^2}{2} \cosh \rho \, \dd t\wedge \dd \rho   \ , \label{Bfield}
\end{equation}
into a Lorentz square term\footnote{The B-field is needed because the divergent term $\tau_{\mu\nu}$ is not positive definite. In the Carrollian limit, where the divergent term is positive definite, there is no need to turn on a B-field, and the rewriting in terms of Lagrange multipliers comes directly.}, 
\begin{equation}
\label{action_F2}
c^2 \gamma^{\alpha\beta} \partial_{\alpha} X^{\mu} \partial_{\beta} X^{\nu} \tau_{\mu\nu}+\varepsilon^{\alpha\beta}  \partial_{\alpha} X^{\mu} \partial_{\beta} X^{\nu} B_{\mu\nu}= c^2 \gamma^{00} \mathcal{F}^A \mathcal{F}^B \tilde{\eta}_{AB}\ ,
\end{equation}
where
\begin{equation}
\label{F}
 \mathcal{F}^A = \tau_{\mu}{}^A \partial_0 X^{\mu} - \frac{1}{\gamma_{11}} \varepsilon^{AB} \tilde{\eta}_{BC} \tau_{\mu}{}^C \partial_1 X^{\mu} - \frac{\gamma_{01}}{\gamma_{11}} \tau_{\mu}{}^A \partial_1 X^{\mu} \ . 
\end{equation}
A quadratic term of this form can be traded off for two Lagrange multipliers as follows
\begin{equation}
\label{rewriting}
c^2 \int \dd^2 \sigma  \, \gamma^{00} \mathcal{F}^A \mathcal{F}^B \tilde{\eta}_{AB} =\int \dd^2 \sigma \bigg(  \lambda_A \mathcal{F}^A - \frac{1}{4 c^2 \gamma^{00}} \lambda_A \lambda^A\bigg) \ .
\end{equation}
Notice that this equivalence only holds on-shell, i.e. solving the equations of motion for $\lambda_A$ and substituting the solution inside the r.h.s. of (\ref{rewriting}) gives back the l.h.s. of that equation. The advantage of the rewriting (\ref{rewriting}) is to trade a divergent term for a finite one, at the price of introducing extra non-dynamical degrees of freedom $\lambda_A$.    
At this point we are finally allowed to take the limit $c \rightarrow \infty$, which gives us the  NR action
\begin{equation}
\label{NR_action}
S^{NR} = - \frac{\sqrt{\lambda}}{4\pi} \int \dd^2 \sigma \, \bigg( \gamma^{\alpha\beta}\partial_{\alpha} X^{\mu} \partial_{\beta} X^{\nu} H_{\mu\nu} + \lambda_A \mathcal{F}^A \bigg) \ . 
\end{equation}
An alternative form of the above action, proposed in \cite{Bergshoeff:2018yvt}, consists in introducing the zweibein for the world-sheet metric $h_{\alpha\beta} = \theta_{\alpha}{}^{i}\theta_{\beta}{}^{j} \eta_{ij}$, such that (\ref{NR_action}) becomes 
\begin{equation}
\label{NR_action_Eric}
S^{NR} = - \frac{\sqrt{\lambda}}{4\pi} \int \dd^2 \sigma \, \bigg( \gamma^{\alpha\beta}\partial_{\alpha} X^{\mu} \partial_{\beta} X^{\nu} H_{\mu\nu} + \varepsilon^{\alpha\beta} (\lambda_+ \theta_{\alpha}{}^+ \tau_{\mu}{}^+ + \lambda_- \theta_{\alpha}{}^- \tau_{\mu}{}^- )\partial_{\beta}X^{\mu}  \bigg) \ ,
\end{equation}
see Appendix \ref{app:Conventions} for our conventions. 
As commented in \cite{Fontanella:2021btt}, this action is obtained by assuming that the following positivity condition must hold
\begin{equation}
\label{positivity_Zweibeine}
\theta_1{}^+ \theta_0{}^- - \theta_0{}^+\theta_1{}^- \geq 0 \ .
\end{equation}  
If instead of choosing this positivity constraint we would have chosen the quantity $\theta_1{}^+ \theta_0{}^- - \theta_0{}^+\theta_1{}^-$ to be negative, then (\ref{NR_action_Eric}) would be the same but with $\theta_{\alpha}{}^+$ and $\theta_{\alpha}{}^-$ swapped. 

The advantage of writing the NR action in the form (\ref{NR_action_Eric}) is to make a $\mathbb{Z}_2$ symmetry manifest. To show this, one needs to notice that (\ref{NR_action_Eric}) is clearly invariant under the following substitution 
\begin{eqnarray}
    \theta_{\alpha}{}^\pm \rightarrow \theta_{\alpha}{}^\mp \ , \qquad \tau_{\beta}{}^\pm \rightarrow \tau_{\beta}{}^\mp  \ , \qquad \lambda_\pm \rightarrow \lambda_\mp \ ,
\end{eqnarray}
where $\tau_{\beta}{}^A$ is the pull-back of $\tau_{\mu}{}^A$.
By taking the explicit form of our $\tau_{\mu}{}^A$ which is diagonal and even in $\rho$, we can understand the swap of $\tau_{\beta}{}^\pm$ as an inversion of the sign of $\rho$. Explicitly,
\begin{equation}
\tau_{\beta}{}^+=\tau_{\mu}{}^+ \partial_{\beta}X^{\mu} = \tau_{t}{}^0  \partial_{\beta} t + \tau_{\rho}{}^1 \partial_{\beta} \rho \qquad\rightarrow \qquad
 \tau_{t}{}^0  \partial_{\beta} t + \tau_{\rho}{}^1 \partial_{\beta} (-\rho) = \tau_{\mu}{}^- \partial_{\beta}X^{\mu}=\tau_{\beta}{}^- \ . 
\end{equation}
We remark that this identification works because $\tau_{\mu}{}^A$ is diagonal and an even function of $\rho$ in our case. Therefore, our NR action enjoys a $\mathbb{Z}_2$ symmetry which consists in 
\begin{eqnarray}
\label{Z2_symm}
\rho \rightarrow - \rho \ , \qquad
\theta_{\alpha}{}^\pm \rightarrow \theta_{\alpha}{}^\mp \ , \qquad
\lambda_\pm \rightarrow \lambda_\mp \ .
\end{eqnarray}
We should emphasise that, although the two formulations of the NR action are equivalent, (\ref{Z2_symm}) is manifest in (\ref{NR_action_Eric}) but not in (\ref{NR_action}). This symmetry will be useful later where we have solutions for the radial coordinate $\rho$ in $\mathbb{R}$ instead of its range $\mathbb{R}_{\geq 0}$. The $\mathbb{Z}_2$ symmetry will allow us to take $|\rho|$ as the physical solution.

\subsection{Simplest classical solutions: $(\kappa, 0, 0)$ and $(\kappa, 0, \nu)$}

Now we have all the ingredients to start computing the equations of motion of the non-relativistic action in conformal gauge. As the action involves also the zweibein associated to the world-sheet metric, we have an additional $SO(1,1)$ freedom, which we choose to fix as $\theta_\alpha\null^\pm= (-1, \mp 1)$.

The equations of motion for $\lambda_A$ in conformal gauge take the form 
\begin{equation}
    \varepsilon^{\alpha \beta} \theta_{\alpha}{}^\pm \tau_{\mu}{}^\pm \partial_{\beta}X^{\mu}=  (\dot{\rho} \mp \rho' )-(t' \mp \dot{t}) \cosh \rho =0 \ , \label{lambdaconstraints}
\end{equation}
where the dot and prime indicate derivatives with respect to $\tau$ and $\sigma$ respectively. After fixing also the residual gauge freedom left by the conformal gauge fixing, which amounts to a Diff$_+ \oplus$Diff$_-$ symmetry, we can completely determine the evolution of $t$ and $\rho$
\begin{eqnarray}
    t = \kappa \tau \ , \qquad
    \rho = \text{gd}^{-1} (\kappa \sigma) =\arcsinh \left( \tan \left( \kappa \sigma \right) \right) \ ,
\end{eqnarray}
where gd stands for the Gudermannian function. We can check that setting the remaining coordinates to zero solves the remaining equations of motion, as well as the Virasoro constraints. In addition, we have to impose periodicity in $\rho$. However, due to the original angular nature of the coordinate $\rho$, which may be periodic up to winding, and the fact that it appears in the metric only inside hyperbolic sines and cosines, we do not demand periodicity of $\rho$ but periodicity of $\cosh \rho$ and $\sinh \rho$. This imposes $\kappa$ to be an integer. This type of solution is called \emph{static solution}, or $(\kappa, 0, 0)$ solution, and all its Noether charges vanish.

There is a more interesting solution which has non-vanishing energy and linear momentum, and it is 
\begin{equation}
    t=\kappa \tau \ , \qquad \rho = \text{gd}^{-1} (\kappa \sigma) \ , \qquad
    \varphi_1=\nu \tau \ , \qquad
    \lambda_{\pm} = \lambda_{\pm}(\sigma) \ . 
\end{equation}
We should stress that $\varphi_1$ was an angular coordinate before taking the non-relativistic limit, but now it resembles a radial coordinate. The fields $\lambda_{\pm}(\sigma)$ need to satisfy the equations of motion for $t$ and $\rho$, which are 
\begin{eqnarray}
\label{eom_t_rho}
    \dot{\lambda}_+ \mp \dot{\lambda}_-= \kappa \tan (\kappa \sigma) \left( \lambda_+ \pm \lambda_- \right) +\left( \lambda'_+ \pm \lambda'_- \right)  \ , 
\end{eqnarray}
and the Virasoro constraints,
\begin{eqnarray}
  - \frac{\nu^2}{2} \pm \kappa |\sec (\kappa \sigma)| \lambda_{\pm} = 0 \ .  
\end{eqnarray}
This is an overdetermined, but consistent, system of equations. We can solve the Virasoro constraints easily, as they are algebraic equations, giving us
\begin{eqnarray}
    \lambda_{\pm} = \pm \frac{\nu^2 |\cos (\kappa \sigma)|}{2 \kappa} \ ,
\end{eqnarray}
which they also solve the differential equation (\ref{eom_t_rho}). In addition, we also have to impose periodicity in $\sigma$ on the Lagrange multipliers, but in this case it gives the same condition as the one coming from demanding periodicity of $\rho$.  We denote this solution as $(\kappa, 0, \nu)$ and it is the equivalent in polar coordinates to the BMN-like solution presented in \cite{Fontanella:2021btt}.

Finally, the only non-vanishing Noether charges of this solution are the energy $E$ and a linear momentum $J$, which was an angular momentum on the 5-sphere before taking the NR limit, and they are
\begin{align}
    E&\equiv -\int_0^{2\pi} \dd\sigma \frac{\dd\mathcal{L}}{\dd\dot{t}}= \frac{\sqrt{\lambda}}{2}\int_0^{2\pi} \frac{\dd\sigma}{2\pi} (\lambda_+-\lambda_-) \cosh \rho = \frac{\sqrt{\lambda}\nu^2}{2\kappa} \ , \label{energysimplest} \\
    J&\equiv \int_0^{2\pi} \dd\sigma \frac{\dd\mathcal{L}}{\dd\dot{\varphi}_1}=\sqrt{\lambda}\nu \ ,
\end{align}
which allow us to write the dispersion relation, 
\begin{equation}
    E = \frac{J^2}{2\kappa \sqrt{\lambda}} \ . \label{dispersion1}
\end{equation}

\subsection{Spinning solution $(\kappa, \omega, \nu)$} \label{moreinvolved}

Let us now move to a more complex ansatz
\begin{equation}
    t=\kappa \tau \ , \qquad \rho = \text{gd}^{-1} (\kappa \sigma) \ , \qquad
    \beta_1=\omega \tau \ ,  \qquad \varphi_1=\nu \tau \ ,  \qquad
    \lambda_{\pm} = \lambda_{\pm}(\sigma) \ .  \label{eq:more_involved}
\end{equation}
with all the remaining coordinates set to zero. Similarly to $\varphi_1$, $\beta_1$ is also an angular coordinate that becomes a radial coordinate after performing the non-relativistic limit.

The fields $\lambda_{\pm}(\sigma)$ are fixed by solving the equations of motion for $t$ and $\rho$,  
\begin{align*}
    \dot{\lambda}_+ - \dot{\lambda}_-&= \kappa \tan (\kappa \sigma) \left( \lambda_+ + \lambda_- \right) +\left( \lambda'_+ + \lambda'_- \right) \ , \\
    \dot{\lambda}_+ + \dot{\lambda}_-&= \kappa \tan (\kappa \sigma) \left( \lambda_+ - \lambda_- \right) +\left( \lambda'_+ - \lambda'_- \right) -2\omega^2 \tan (\kappa \sigma) |\sec (\kappa \sigma)| \ ,
\end{align*}
and by solving the Virasoro constraints,
\begin{eqnarray}
  - \nu^2 -\omega^2 \tan^2 (\kappa \sigma) \pm 2\kappa |\sec (\kappa \sigma)| \lambda_{\pm} = 0 \ .  
\end{eqnarray}
Again, this system of equations is overdetermined but consistent. It is simpler to solve the Virasoro constraints, as they are algebraic equations for $\lambda_{\pm}$, giving us
\begin{eqnarray}
    \lambda_{\pm} = \pm \frac{\nu^2 + \omega^2 \tan^2 (\kappa \sigma)}{2 \kappa  |\sec (\kappa \sigma)|} \ .
\end{eqnarray}
Similarly, periodicity of $\sinh \rho$, $\cosh \rho$ and $\lambda_\pm$ implies that $\kappa$ is an integer. This solution is denoted by $(\kappa, \omega, \nu)$.

The Noether charges associated to this solution are
\begin{align}
    E&\equiv -\int_0^{2\pi} \dd\sigma \frac{\dd\mathcal{L}}{\dd\dot{t}}= \frac{\sqrt{\lambda}}{4 \pi} \int_0^{2\pi} \dd\sigma  (\lambda_+-\lambda_-) \cosh \rho  \notag\\
    &= \frac{\sqrt{\lambda}}{2}  \left( \frac{\nu^2}{\kappa} + \frac{\omega^2}{\kappa}  \int_0^{2\pi} \frac{\dd\sigma}{2\pi}  \tan^2 (\kappa \sigma) \right)  \ , \label{dispersion2} \\
    S&\equiv \int_0^{2\pi} \dd\sigma \frac{\dd\mathcal{L}}{\dd\dot{\beta}_1}= \sqrt{\lambda} \,\omega \int_0^{2\pi} \frac{\dd\sigma}{2\pi} \tan^2 (\kappa \sigma) \ , \label{SpinInvolved}\\
    J&\equiv \int_0^{2\pi} \dd\sigma \frac{\dd\mathcal{L}}{\dd\dot{\varphi}_1}= \sqrt{\lambda} \nu \ . \label{JInvolved}
\end{align}
Notice that the integral of the square of the tangent diverges, as it introduces a second order pole in the integration path. Thus, the only solution with finite energy is the one with $\omega=0$, namely, the solution we studied in the previous section. However, there exist another interesting case, which corresponds to $\omega=\pm \kappa$. In this case, energy and spin are both divergent, but their linear combination is finite, that is
\begin{equation}
   E \mp\frac{S}{2}=\frac{J^2}{2\kappa \sqrt{\lambda}} \ . \label{dispersion3}
\end{equation}

\section{Relativistic strings flowing to their NR counterparts}
\label{sec:flow_rel}

In this section, we take a step back and approach the construction of NR classical string solutions from a different angle. Instead of considering the NR limit at the level of the action, we want now to keep the contraction parameter $c$ finite, and take the large $c$ limit only at the very end. We start with the relativistic action (\ref{rel_action}), which is coupled to the critical closed B-field, and we rescale coordinates with $c$ accordingly to (\ref{coord_resc}). If $c=1$, we get the relativistic action, but if $c \neq 1$, we have a theory that interpolates between the relativistic and the NR actions.\footnote{As long as $c$ is finite, the action is equivalent to the relativistic one because we can always integrate out the Lagrange multiplies using eq.~\eqref{rewriting}. However, at large values of $c$ one should be able to construct an interpolating theory in the spirit of \cite{Hartong:2021ekg, Hartong:2022dsx}.} We solve the $c$-dependent equations of motion in conformal gauge, obtaining solutions depending on $c$, we take the $c\rightarrow \infty$ limit on the classical solutions, and compare with the NR strings from the previous section.

Here we will consider two particular examples: the BMN string and a non-compact version of the folded string. We will show that there is no physically meaningful NR string associated to the first one, and that the second one, in both cases with and without spin, reduces to the classical NR strings discussed in the previous section.

\subsection{Flow of BMN string}

The two simplest solutions to the equations of motion for strings propagating in AdS$_5\times$S$^5$ are the massless geodesic, which are point-like strings. They are classified, up to global $SO(2,4)\times SO(6)$ transformations, into two types: either the geodesic lies completely inside AdS$_5$, or time evolves in AdS$_5$ and the string moves around the big circle of S$^5$. In the latter case, the solution is the famous BMN. 

In the coordinates used in (\ref{coord}), the BMN string takes the form
\begin{equation}
    t= \kappa \tau \ , \qquad \varphi_1=c\kappa \tau \ .
\end{equation}
with all the other coordinates set to zero. We can see that the large $c$ limit is not well-defined here. Although $\varphi_1$ is an angular variable defined modulo $2\pi$, trigonometric functions acting on $\varphi_1$ will be ill-defined. Energy and angular momentum associated with this solution are
\begin{align}
\label{BMN_E_J}
    E\equiv -\int_0^{2\pi} \dd\sigma \frac{\dd\mathcal{L}}{\dd\dot{t}}= \sqrt{\lambda} c^2 \kappa \ , \qquad
    J\equiv \int_0^{2\pi} \dd\sigma \frac{\dd\mathcal{L}}{\dd\dot{\varphi}_1}= \sqrt{\lambda} c \kappa \ , \qquad 
    E= cJ \ , 
\end{align}
which become infinite in the large $c$ limit.  Equivalently, we can argue that $\kappa$ should go to zero as $c$ goes to infinity, such that $\tilde{\kappa}=c\kappa$ remains finite. In this picture, we obtain an unphysical solution localised at $t=0$ with finite $J$ but still infinite $E$. Thus, we can argue that we cannot define a physically meaningful non-relativistic limit of the BMN solution.

From a physical perspective, non-relativistic string vacua should have a winding along the spatial longitudinal direction in order to have a well-defined spectrum. However, the BMN solution is point-like both in AdS$_5$ and S$^5$ spaces, and therefore cannot have winding. It is possible to generalise such BMN solution by adding winding on a circle inside S$^5$. However, such winding is not aligned with the critical Kalb-Ramond B-field defined in (\ref{Bfield}).

\subsection{Flow of non-compact folded string}

The folded string solution is one of the most famous classical string solutions in AdS$_5 \times $S$^5$. However, for our purposes, we need to consider a non-compact version of the usual folded string.
Both, the usual and the non-compact folded strings are characterised by the ansatz
\begin{equation}
    t=\kappa \tau \ , \quad \beta_1=\omega \tau \ , \quad \varphi_1=\nu \tau \ , \quad \rho=\rho (\sigma) \ . \label{foldedansatz}
\end{equation}
All the remaining coordinates are set to zero. If we substitute this ansatz into the equations of motions, the only equation that is non-trivially satisfied is
\begin{equation}
    c^2\rho''=(c^2\kappa^2 -\omega^2) \sinh \rho \cosh \rho \ , \label{eq:EoMFolded}
\end{equation}
which should be supplemented with the Virasoro constraint
\begin{equation}
    c^2\rho^{\prime 2}=(c^2\kappa^2 - \nu^2) \cosh^2 \rho - (\omega^2 - \nu^2) \sinh^2 \rho \ . \label{eq:VirasoroFolded}
\end{equation}
The solution to these equations is unique and given by\footnote{We thank Arkady Tseytlin for remarking us that the range of $\rho$ is defined from $0$ to $+ \infty$, as $\sinh \rho$ is a radial coordinate. Thanks to the $\mathbb{Z}_2: \rho \rightarrow -\rho$ symmetry on the relativistic metric, $|\rho |$ is also a solution of the equations of motion, which belongs to the appropriate range. However, we will not consider $|\rho |$ in the intermediate steps because this introduces an apparent singularity in the computations, but we will keep it in mind in final results.}
\begin{eqnarray}
\label{rho_sol}
    \rho = - i \text{ am} \left( \sqrt{ \frac{\nu^2}{c^2} - \kappa^2 } \sigma , \frac{c^2 \kappa^2 - \omega^2}{c^2 \kappa^2 - \nu^2} \right)
\end{eqnarray}
In the regime $\nu > c\, \kappa$, $\sinh \rho$ is a compact function and we find the usual folded string, see e.g. \cite{Gubser:2002tv,Frolov:2002av, Frolov:2006qe, Beccaria:2008dq}. However, when $\nu < c \,\kappa$, $\sinh \rho$ is a non-compact function given by\footnote{In this article we will follow the convention of \cite{Abramowitz} for the elliptic modulus, where dn$(x,m) + m$ sn$(x,m)=1$ is fulfilled.}  
\begin{align}
\label{noncompactsol}
    \sinh \rho &= -\sqrt{\frac{c^2\kappa^2 - \nu^2}{c^2\kappa^2 - \omega^2 }}\text{sc} \left(\frac{\sqrt{c^2\kappa^2 - \omega^2 }}{c}  \sigma , \frac{\nu^2 -\omega^2}{c^2\kappa^2 - \omega^2 }\right) \ , 
\end{align}
where we have assumed that $\nu^2 >\omega^2$.\footnote{This non-compact solution can also be obtained from the usual compact folded string solution using the identity $\text{sn} (ix,m)=i\, \text{sc} (x,1-m)$. This identity allows us to show that the large $c$ limit of the compact folded string also flows to (\ref{foldedansatz_large_c}).} We demand that the above solution must be periodic in the $\sigma$ coordinate. The condition $\sinh \rho(\sigma + 2\pi) = \sinh \rho (\sigma)$ imposes 
\begin{eqnarray}
\label{periodicity}
    \frac{\pi}{2}  \sqrt{\frac{c^2\kappa^2 - \omega^2 }{c^2}} = n \, \text{K}\left( \frac{\nu^2 -\omega^2}{c^2\kappa^2 - \omega^2 } \right)
\end{eqnarray}
where $n$ is an integer. The large $c$ limit of (\ref{foldedansatz}) gives 
\begin{equation}
    t=\kappa \tau \ , \quad \beta_1=\omega \tau \ , \quad \varphi_1=\nu \tau \ , \quad \rho=-i\, \text{gd} ( i\kappa \sigma)=\text{gd}^{-1} ( \kappa \sigma) \ , \label{foldedansatz_large_c}
\end{equation}
while the large $c$ limit of (\ref{periodicity}) gives
\begin{equation}
    \frac{\pi}{2} |\kappa|=n \, \text{K}(0) \ \ \Longrightarrow \ \ \kappa\in \mathbb{Z} \, ,
\end{equation}
which matches perfectly the more involved NR classical solution (\ref{eq:more_involved}) (without the Lagrange multipliers part, as we do not have access to them coming from this perspective). Thus, the non-compact folded string is the correct relativistic origin of the NR classical solutions we studied in the previous section.

We shall show now that the dispersion relation of the non-compact folded string also gives rise to the dispersion relations we found for the NR solutions. In contrast with the usual folded string, this non-compact version has not been widely studied in the literature because its spin and energy diverge, which makes this analysis a bit more involved. However, as we are motivated by NR string theory, the construction of the SNC AdS$_5 \times$S$^5$ string action requires us to include a B-field that provides an additional contribution to the classical energy. We will study separately the cases of $\omega=0$ and $\omega \neq 0$ because in the first case the B-field contribution is enough to cancel such divergence.

\subsubsection{Solution $(\kappa, 0, \nu)$}

Let us address first the case of $\omega=0$. As we have already stated, the energy has two Lagrangian contributions: one coming from the AdS$_5 \times$S$^5$ metric, $E^G$, and one from the closed B-field  (\ref{Bfield}), $E^B$,\footnote{As we commented above, we have to restrict $\rho$ to non-negative values. This amount to substituting $\rho$ by $|\rho|$ in the following expressions. Nevertheless, the discussion on the cancellation of the divergence remains the same. We thank Andrea Guerrieri for useful comments on this point.} 
\begin{align}
    E^G &\equiv -\int_0^{2\pi} \dd\sigma \frac{\dd\mathcal{L}^G}{\dd\dot{t}}= c^2 \sqrt{\lambda} \int_0^{2\pi} \frac{\dd\sigma}{2\pi} \cosh^2\rho \, \dot{t} \ , \\
    E^B &\equiv -\int_0^{2\pi} \dd\sigma \frac{\dd\mathcal{L}^B}{\dd\dot{t}} = -c^2 \sqrt{\lambda} \int_0^{2\pi} \frac{\dd\sigma}{2\pi} \cosh\rho \, \rho' \ ,\\
    E &= E^G + E^B = c^2 \sqrt{\lambda} \int_0^{2\pi} \frac{\dd\sigma}{2\pi} \cosh\rho \, (\cosh\rho \, \dot{t}-\rho') \ . \label{renormalisedenergy}
\end{align}
For any non-vanishing value of $c$, $E^G$ is divergent due to the non-compactness of $\rho$. This manifest as second order poles in the integration path. One is located at  $\sigma = \frac{\pi}{2}$,
\begin{equation}
    \cosh^2\rho \, \dot{t}= 
    \kappa \, \text{dc}^2 \left( \kappa \sigma , \frac{\nu^2}{c^2 \kappa^2} \right) \sim \frac{1}{\kappa (\sigma - \pi/2)^2} \ ,
\end{equation}
and another one is located at $\sigma = \frac{3\pi}{2}$ with the same residue. However, this is also true for $E^B$ but with opposite residue
\begin{equation}
    -\cosh\rho \, \rho' = 
    -\sqrt{\kappa^2 - \nu^2} \, \text{dc} \left( \kappa \sigma , \frac{\nu^2}{c^2 \kappa^2} \right) \text{nc} \left( \kappa \sigma , \frac{\nu^2}{c^2 \kappa^2} \right) \sim -\frac{1}{\kappa  (\sigma - \pi/2)^2} \ ,
\end{equation}
and similarly for the one at $3\pi/2$. Thus, the combined contribution to the energy is finite and given by
\begin{eqnarray}
\label{energy_omega_0}
    E = \frac{2nc^2 \sqrt{\lambda}}{\pi} \left[ \text{K} \left( \frac{\nu^2}{c^2 \kappa^2} \right) - \text{E} \left( \frac{\nu^2}{c^2 \kappa^2} \right) \right]=\kappa c^2 \sqrt{\lambda} \left[ 1 - \frac{\text{E} \left( \frac{\nu^2}{c^2 \kappa^2} \right)}{\text{K} \left( \frac{\nu^2}{c^2 \kappa^2} \right)} \right] \ ,
\end{eqnarray}
where $n$ is the integer appearing in the periodicity condition (\ref{periodicity}). This solution has also an angular momentum $J$ 
\begin{equation}
\label{J_omega_0}
    J \equiv \int_0^{2\pi} \dd\sigma \frac{\dd\mathcal{L}}{\dd\dot{\varphi}_1}=\sqrt{\lambda}\nu \ . 
\end{equation}
We cannot write a closed form for the dispersion relation because it requires us to solve the periodicity condition, which cannot be solved analytically due to the elliptic integral.

In the large $c$ limit, the energy (\ref{energy_omega_0}) is still finite\footnote{We should point out that the energy (\ref{renormalisedenergy}) is proportional to the difference of the two constrains imposed by the Lagrange multipliers in SNC AdS$_5 \times$S$^5$ (\ref{lambdaconstraints}). This assures us that the limit is finite despite the overall $c^2$ factor.} and becomes
\begin{equation}
    \lim_{c\rightarrow \infty} E= \lim_{c\rightarrow \infty} \kappa c^2 \sqrt{\lambda} \left[ 1 - \frac{\text{E} \left( \frac{\nu^2}{c^2 \kappa^2} \right)}{\text{K} \left( \frac{\nu^2}{c^2 \kappa^2} \right)} \right] = \frac{\sqrt{\lambda}\nu^2}{2 \kappa} \ ,
\end{equation}
which matches perfectly the energy (\ref{energysimplest}). The angular momentum $J$ will remain the same in the large $c$ limit, and therefore we recover the dispersion relation (\ref{dispersion1}).

\subsubsection{Spinning solution $(\kappa, \omega, \nu)$}

In this section, we consider the case $\omega \neq 0$. The analysis is similar to the $\omega=0$ case, so we will only point out the differences. The energy again gets a contribution from the metric and the B-field, as defined in (\ref{renormalisedenergy}). However, in this case, the poles that appear in the contribution to the energy from the B-field do not cancel the ones that appear in the contribution from the metric. In fact, 
\begin{equation}
\cosh^2 \rho \, \dot{t} - \cosh \rho \, \rho' \sim \frac{\kappa - \sqrt{\kappa^2 - \frac{\omega^2}{c^2}}}{\kappa^2 - \frac{\omega^2}{c^2}} \frac{1}{(\sigma - \pi / 2)^2} \ ,
\end{equation}
and similarly for $\sigma=3\pi/2$.

Despite that, we can still formally study what is the large $c$ limit of the energy. After some non-trivial algebra, we find
\begin{equation}
   \lim_{c\rightarrow \infty} E =  \frac{ \kappa  \sqrt{\lambda}}{2\pi} \int_0^{2\pi} \frac{\nu^2 + \omega^2 \tan^2 (\kappa \sigma) }{2 \kappa^2} \dd \sigma \ .
\end{equation}
It is immediate to see that the resulting energy is exactly the same as (\ref{dispersion2}), that is, the energy of the NR classical solution described in section~\ref{moreinvolved}.

The spin $S$ is given by
\begin{equation}
    S\equiv \int_0^{2\pi} \dd \sigma \frac{\dd \mathcal{L}}{\dd \dot{\beta}_1} =\sqrt{\lambda} \omega \int_0^{2\pi} \frac{\dd\sigma}{2\pi} \sinh^2\rho \ .
\end{equation}
If we substitute formula (\ref{noncompactsol}) in this equation and compute its large $c$ limit, we obtain 
\begin{equation}
    \lim_{c\rightarrow \infty} S =\sqrt{\lambda} \omega \int_0^{2\pi} \frac{\dd\sigma}{2\pi} \tan^2 (\kappa \sigma ) \ ,
\end{equation}
which matches (\ref{SpinInvolved}) perfectly.

For completeness, we should mention that the angular momentum $J$ for this solution is also $\sqrt{\lambda} \nu$, which matches (\ref{JInvolved}).

At the end, after fixing $\omega = \pm \kappa$, we reproduce precisely the dispersion relation (\ref{dispersion3}).\footnote{We need to be careful when deriving this result, as choosing $\omega = \pm \kappa$ is not consistent at $c=1$ since $\sinh(\rho)$ becomes purely imaginary. This can be cured by taking $\kappa = \pm (\omega - 1/c)$ at the relativistic level.}  
Therefore, we can safely conclude that the NR classical strings in SNC AdS$_5 \times$S$^5$ described in the previous section can be reconstructed from the large $c$ limit of the non-compact folded string in AdS$_5 \times$S$^5$.

\subsection{A comment on the non-compact folded string in Cartesian coordinates}
\label{sec:comment_cartesian}

We want to close this section with a comment on the large $c$ limit of the non-compact folded string in Cartesian coordinates. At the relativistic level, it is physically equivalent to write the AdS$_5 \times$S$^5$ string theory action in different sets of coordinates, in particular, there is no problem in changing from polar coordinates to Cartesian ones. This stops to hold after taking the non-relativistic limit, and one expects to obtain equivalent theories only when the change of coordinates transformation is analytic in $1/c^2$.\footnote{ For a further discussion on this topic, see also \cite{Hansen:2019vqf, Fontanella:2021hcb}.} This is not our case, as the change of coordinates is non-invertible at $\rho=0$. Because of that, we want to analyse separately the non-relativistic limit of AdS$_5 \times$S$^5$ in Cartesian coordinates.

Although the actions may be different in the NR limit, the relativistic classical string solution that acts as seed is the same, so we may be able to change to Cartesian coordinates and apply the same logic to reconstruct the solutions presented in \cite{Fontanella:2021btt}. For that, we consider AdS$_5$ metric in Cartesian coordinates 
\begin{equation}
\label{AdS_metric_cartesian}
\dd s^2_{\text{AdS}} = - \bigg(\frac{1+ \frac{z_i z^i}{4}}{1-\frac{z_i z^i}{4}}\bigg)^2 \dd t^2 + \frac{1}{(1-\frac{z_i z^i}{4})^2} \dd z_i \dd z^i \ , \qquad\qquad i = 1, ..., 4\ .
\end{equation}
The diffeomorphism that takes the Cartesian metric (\ref{AdS_metric_cartesian}) to the polar one (\ref{coord}) is 
\begin{gather*}
    z_1=2 \tanh \left(\frac{\rho}{2} \right) \, \cos \beta_3 \, \cos \beta_2 \, \cos \beta_1 \ , \qquad z_2=2 \tanh \left(\frac{\rho}{2} \right) \, \cos \beta_3 \, \cos \beta_2 \, \sin \beta_1 \ , \\
    z_3=2\tanh \left(\frac{\rho}{2} \right) \, \cos \beta_3 \, \sin \beta_2 \ , \qquad z_4=2\tanh \left(\frac{\rho}{2} \right) \, \sin \beta_3 \ , \qquad t=t \ .
\end{gather*}
Substituting our ansatz for the folded string (\ref{foldedansatz}), we get
\begin{equation}
    t=\kappa \tau \ , \qquad z_1= 2\tanh \left(\frac{\rho}{2} \right) \, \cos (\omega \tau) \ , \qquad z_2= 2\tanh \left(\frac{\rho}{2} \right) \, \sin (\omega \tau) \ .
\end{equation}
Considering $\omega=0$ and the large $c$ limit, we have
\begin{equation}
    t=\kappa \tau \ , \qquad z_1= 2\tan \left(\frac{\kappa \sigma}{2} \right)  \ ,
\end{equation}
where we have used that the inverse Gudermannian function can also be expressed as gd$^{-1}(x)=2\arctanh (\tan (x/2))$. From that, it is clear that the BMN-like string we found in \cite{Fontanella:2021btt} is exactly the non-relativistic limit of the non-compact folded string presented here in polar coordinates.

\section{Conclusions}
\label{sec:conclusions}

Following the line of research started in \cite{Fontanella:2021btt} in Cartesian coordinates, we have constructed the simplest classical string solutions of NR string theory in SNC AdS$_5 \times$S$^5$ in polar coordinates. 
In this article, we answer the question of what is the relativistic origin of these NR string solutions. We found that the relativistic string that at large $c$ becomes the BMN-like string of NR string theory in SNC AdS$_5 \times$S$^5$ is a non-compact version of the folded string with zero spin $S$. Such solution has infinite energy and, due to that, it has been ignored so far in the literature. However, motivated by NR string theory, we need to couple the original relativistic action in AdS$_5 \times$S$^5$ to a closed critical B-field. This B-field does not contribute to the equations of motion, but it modifies the energy of the string and, in fact, its contribution to the energy has the precise form to cancel the divergence of the energy coming from the metric contribution. Such cancellation holds at each value of the parameter $c$. In the limit when $c$ is large, we found that the total energy (metric and B-field) is still finite and matches precisely the energy of the BMN-like string computed by using the NR string action.

In addition to the BMN-like string, we also considered a more complicated NR string with spin $S$ in the AdS transverse directions. This string has infinite energy and spin but, for some particular values of the free parameters ($\kappa = \pm \omega$), their difference is finite. Again, we found the relativistic solution that flows to it in the large $c$ limit. Such solution is a non-compact version of the folded string with spin $S$. This solution also has infinite energy and spin, and this time the B-field contribution is not enough to cancel the divergences. In contrast to the non-compact folded string with zero spin $S$, this solution has been considered in the literature, e.g. \cite{Gubser:2002tv}, although not in much detail. In the limit when $c$ is large, we found that the total energy matches the one of the string solution computed by using the NR string action. 

Interestingly, we found that the large $c$ limit of the BMN string does not lead to any consistent NR classical string. This seems to be related with the observation from \cite{Gomis:2000bd} in flat space, which states that the spectrum of strings with no winding is empty, as the BMN string is point-like. The NR limit represents a way to zoom into a special corner of the relativistic theory, where only strings with slow velocity can appear. In line with this picture, it seems like the BMN string is too fast to be seen at the NR regime. 

The ideas of this article could be applied to more exotic relativistic classical string solutions, e.g. spiky string \cite{Kruczenski:2004wg,Ryang:2005yd}, pulsating string \cite{Minahan:2002rc,Engquist:2003rn,Dimov:2004xi,Smedback:2004udl}, giant magnons \cite{Hofman:2006xt,Dorey:2006dq,Chen:2006gea}, etc.   
As there is already a vast literature on classical strings in AdS$_5 \times$S$^5$, this allows us to borrow those results instead of needing to perform a classification from scratch. Then it would be interesting to study the large $c$ flow of all of them and see which one remains in the NR corner. It would also be interesting to see if the Pohlmeyer reduction \cite{Grigoriev:2007bu,Miramontes:2008wt,Hoare:2009rq} can be applied in this limit and if there exist a NR version of the Neumann and Neumann-Rosochatius integrable system that appears in the context of spinning strings in AdS$_5 \times$S$^5$ \cite{Arutyunov:2003uj,Arutyunov:2003za,Kruczenski:2006pk}.

The NR string dispersion relation for flat space was obtained in \cite{Gomis:2000bd} by taking the zero Regge limit of the relativistic one. In flat space, computing the spectrum of fluctuations is a vacua independent result, as the action is Gaussian, which is not the case in AdS$_5 \times$S$^5$. Here, we have shown that the classical part of the spectrum, namely the dispersion relation of the classical solution, can be obtained from the large $c$ limit of the relativistic one. It is still an open question if this picture survives at the quantum level. This is not easy to check, as we do know neither the quantum corrections of the NR classical solutions nor of the associated relativistic ones. Although the quantum corrections of the relativistic (compact) folded string are well known \cite{Frolov:2002av,Beccaria:2010ry}, it is not obvious if those results immediately extend to the non-compact string we have used in this article. Even more, it is not even clear to us if the fluctuations around this non-compact folded string are well-defined, as only the solution with $S=0$ has finite total energy once the B-field is included in the action. Regarding the NR classical strings, some attempts have been made to access the quantum corrections \cite{Fontanella:2021hcb, Fontanella:2022wfj}. These articles use well-known methods employed in relativistic AdS$_5 \times$S$^5$, the light-cone quantisation and the classical spectral curve, but it has been found that they do not completely work for the NR action. The results presented here may shed some light on how to adapt these methods to SNC backgrounds.

From the holographic point of view, it would be interesting to understand the dual role of the closed critical B-field in $\mathcal{N}=4$ SYM, as it should turn on a Tr$F^3$ term \cite{Das:1998ei, Ferrara:1998bp}\footnote{We thank Elias Kiritsis for pointing out these references to us.}. It might be that the bare conformal dimension of the dual gauge invariant operator corresponding to the non-compact folded string with $S=0$ becomes infinite (infinite spin chain) and the Tr$F^3$ term might act as a counterterm that makes it finite.

\section*{Acknowledgments}

We thank Andrea Guerrieri, Marius de Leeuw,  Arkady Tseytlin, Pedro Vieira for useful discussions and Arkady Tseytlin for valuable feedback on a draft of this work. We thank the participants of the workshop \emph{Non-Relativistic Strings and Beyond} at Nordita for many stimulating discussions and useful feedback on our work. Research at Perimeter Institute is supported in part by the Government of Canada through the Department of Innovation, Science and Economic Development and by the Province of Ontario through the Ministry of Colleges and Universities. During the final part of this work, AF was supported by the IRN:QFS scheme (International Research Network on Quantum Fields and Strings) to visit Nordita. JMNG is supported by the Deutsche Forschungsgemeinschaft (DFG, German Research Foundation) under Germany's Excellence Strategy -- EXC 2121 ``Quantum Universe'' -- 390833306.
AF thanks Lia for her permanent support.

\begin{appendices}

\section{Conventions}
\label{app:Conventions}

For a generic object $\mathcal{O}^A$, we define its light-cone combinations as
\begin{equation}\label{LC_comb}
\mathcal{O}^{\pm} \equiv \mathcal{O}^0 \pm \mathcal{O}^1 \ , \qquad\qquad
\mathcal{O}_{\pm} \equiv \frac{1}{2}\left( \mathcal{O}_0 \pm \mathcal{O}_1 \right) \ .
\end{equation}
The longitudinal Minkowski metric then has non-vanishing components $\eta_{+-} = -1/2$ and $\eta^{+-} = -2$. We take $\varepsilon^{01} = - \varepsilon_{01} = + 1$ for $\varepsilon^{\alpha\beta}$, $\varepsilon^{\sf ab}$ and $\varepsilon^{AB}$. In light-cone components $\varepsilon_{+-} = \frac{1}{2}$, $\varepsilon^{+-} = -2$.
Our convention for $p$-forms is $\omega_p = \frac{1}{p!} \omega_{\mu_1 \cdots \mu_p} \dd x^{\mu_1}\wedge \cdots \wedge \dd x^{\mu_p}$.

\section{Polar coordinates coset representative and NR limit}
\label{App:polar_coset}

The AdS$_5 \times$S$^5$ metric in polar coordinates (\ref{coord}) can be written in terms of a Maurer-Cartan (MC) 1-form, since AdS$_5 \times$S$^5$ is the coset space $SO(2,4) \times SO(6) / SO(1,4) \times SO(5)$. The $\mathfrak{so}(2,4) \oplus \mathfrak{so}(6)$ algebra is generated by relativistic translations $P_{\hat{a}}$ and rotations $J_{\hat{a}\hat{b}}$ for the AdS$_5$ part, with $\hat{a}, \hat{b}, ... = 0, 1, ..., 4$, whereas for the S$^5$ part it is generated by spatial translations $\text{P}_{a'}$ and rotations $\text{J}_{a'b'}$, with $a', b', ... = 1, ..., 5$. Its commutation relations are
\begin{subequations}\label{so(4,2)+so(6)}
	\begin{align}
	[P_{\hat{a}}, P_{\hat{b}}] &= \frac{1}{R^2} J_{\hat{a}\hat{b}} \ , &
	[\text{P}_{a'}, \text{P}_{b'}] &= - \frac{1}{R^2} \text{J}_{a'b'} \ , \\
	[P_{\hat{a}}, J_{\hat{b}\hat{c}}] &= 2 \eta_{\hat{a}[\hat{b}} P_{\hat{c}]} \ , &
	[\text{P}_{a'}, \text{J}_{b'c'}] &= 2 \delta_{a'[b'} \text{P}_{c']} \ , \\
	[J_{\hat{a}\hat{b}}, J_{\hat{c}\hat{d}}] &= 4\eta_{[\hat{b}[\hat{c}} J_{\hat{a}]\hat{d}]} \ , &
	[\text{J}_{a'b'}, \text{J}_{c'd'}] &= 4\delta_{[b'[c'} \text{J}_{a']d']} \ .
	\end{align}
\end{subequations}
where $R$ is the common AdS$_5$ and S$^5$ radius. 
The algebra has a $\mathbb{Z}_2$ outer automorphism, where the $\{J_{\hat{a}\hat{b}}, \text{J}_{a'b'}\}$ span a grading 0 subspace, whereas the complementary set $\{ P_{\hat{a}}, \text{P}_{a'} \}$ span a grading 1 subspace. One can construct a MC 1-form $A = g^{-1} \dd g$, with $g\in \mathfrak{so}(2,4) \oplus \mathfrak{so}(6)$, which takes the form 
\begin{eqnarray}
    A_{\mu} = e_{\mu}{}^{\hat{a}} P_{\hat{a}} + e_{\mu}{}^{a'} \text{P}_{a'} + \omega_{\mu}{}^{\hat{a}\hat{b}} J_{\hat{a}\hat{b}} + \omega_{\mu}{}^{a'b'}\text{J}_{a'b'} \ ,
\end{eqnarray}
where $e_{\mu}{}^{\hat{a}}, e_{\mu}{}^{a'}$ are the vielbein of AdS$_5 \times$S$^5$ and $\omega_{\mu}{}^{\hat{a}\hat{b}}, \omega_{\mu}{}^{a'b'}$ are the components of the Levi-Civita spin connection.  
The metric is then obtained by taking the ``square'' of an MC 1-form,  
\begin{eqnarray}
    g_{\mu\nu} = \langle A_{\mu}^{(1)} , A_{\nu}^{(1)} \rangle \ ,  
\end{eqnarray}
where $A_{\mu}{}^{(1)}$ is the projection of $A_{\mu}$ inside the grading 1 subspace, and $\langle \cdot, \cdot \rangle$ is an inner product, adjoint invariant under $\mathfrak{so}(2,4) \oplus \mathfrak{so}(6)$, which is taken to be 
\begin{eqnarray}
    \langle P_{\hat{a}}, P_{\hat{b}} \rangle = \eta_{\hat{a}\hat{b}} \ , \qquad
     \langle \text{P}_{a'}, \text{P}_{b'} \rangle = \delta_{a'b'} \ . 
\end{eqnarray}
The AdS$_5 \times$S$^5$ metric in polar coordinates given in (\ref{coord}) can be written in terms of the following choice of coset representative, which we have not found in literature and we are presenting here for the first time,
\begin{align}
\label{polar_rep}
    g &= g_{\text{AdS}} g_{\text{S}} \ , \\
    \notag
    g_{\text{AdS}} &= \exp(t P_0) \exp(\beta_3 J_{34}) \exp\left((\beta_2 + \frac{\pi}{2}) J_{13}\right) \exp(\beta_1 J_{12}) \exp(\rho P_1) \\
    \notag
    g_{\text{S}} &= \exp(\varphi_5 \text{J}_{34}) \exp\left( (\varphi_4 + \frac{\pi}{2}) \text{J}_{13} \right) \exp(\varphi_3 \text{J}_{12}) \exp\left( (\varphi_2 + \frac{\pi}{2}) \text{P}_1\right) \exp(\varphi_1 \text{P}_5 ) \ , 
\end{align}
where $\{ t, \rho, \beta_1, \beta_2, \beta_3 \}$ are coordinates of AdS$_5$, whereas $\{ \varphi_1, ..., \varphi_5 \}$ are coordinates of S$^5$. 

The NR limit proposed in \cite{Gomis:2005pg} consists in taking the \.In\"on\"u-Wigner contraction of $\mathfrak{so}(2,4) \oplus \mathfrak{so}(6)$ to the string Newton-Hooke$_5 \oplus$Eucl$_5$ algebra. This amounts of splitting the AdS indices as $\hat{a} = (A, a)$, with $A=0,1$ and $a=2, 3,4$, and rescaling the generators, as well as the radius $R$, as 
\begin{eqnarray}
    P_A \rightarrow \frac{1}{c} P_A \ , \qquad
    J_{Aa} \rightarrow c\, J_{Aa} \ , \qquad
    R \rightarrow c R \ ,
\end{eqnarray}
which is equivalent to \emph{not} rescaling the radius $R$, but acting only on the generators,
\begin{eqnarray}
\label{resc_generators}
    P_a \rightarrow c\, P_a \ , \qquad
    \text{P}_{a'} \rightarrow c \,\text{P}_{a'} \ , \qquad
    J_{Aa} \rightarrow c \, J_{Aa} \ ,
\end{eqnarray}
and then taking the $c \to \infty$ limit. Thanks to the choice of coset representative (\ref{polar_rep}), generators are in a 1:1 correspondence with coordinates. One can then pass to the ``dual'' picture, where the algebra is not rescaled, but coordinates are. In this dual picture, the rescaling of generators (\ref{resc_generators}) is equivalent to rescaling coordinates, as
\begin{equation}
\beta_2 + \frac{\pi}{2} \rightarrow \frac{1}{c} \big(\beta_2 + \frac{\pi}{2} \big) \ , \qquad
\beta_1 \rightarrow \frac{1}{c} \beta_1 \ , \qquad
\varphi_2 + \frac{\pi}{2} \rightarrow \frac{1}{c} \big(\varphi_2 + \frac{\pi}{2} \big) \ , \qquad
\varphi_1 \rightarrow \frac{1}{c} \varphi_1 \ .
\end{equation}
The advantage of having used the MC formalism is that we are not forced to use the same coordinates as in \cite{Gomis:2005pg} to define the NR limit. In this way, the NR limit is defined as a contraction of the isometry algebra, which in turns fixes the rescaling of one's own favourite coordinates, as done in \cite{Fontanella:2021hcb}. One should keep in mind that different choices of coordinates can land to different NR theories.


\end{appendices}


\bibliographystyle{nb}

\bibliography{Biblio.bib}

\end{document}